\documentclass[12pt]{article}

\usepackage{amsmath}
\usepackage{amsfonts}        
\usepackage{amssymb}
\usepackage{amsbsy}
\usepackage{mathrsfs}
\usepackage{graphicx}
\usepackage{dcolumn}
\usepackage{bm}
\usepackage{xcolor}
\usepackage{hyperref}
\usepackage{cancel}

\begin{document}

\newtheorem{theorem}{Theorem}
\newtheorem{definition}{Definition}
\newtheorem{lemma}{Lemma}
\newtheorem{proposition}{Proposition}
\newtheorem{remark}{Remark}
\newtheorem{corollary}{Corollary}
\newtheorem{example}{Example}

\newcommand{\BEQ}{\begin{equation}}     
\newcommand{\BEA}{\begin{eqnarray}}
\newcommand{\BD}{\begin{displaymath}}
\newcommand{\EEQ}{\end{equation}}       
\newcommand{\EEA}{\end{eqnarray}}
\newcommand{\ED}{\end{displaymath}}
\newcommand{\e}{{\rm e}}
\newcommand{\pr}{{\rm Pr}}
\newcommand{\eps}{\varepsilon}          
\newcommand{\vep}{\varepsilon}          
\newcommand{\vph}{\varphi}              
\newcommand{\vro}{\varrho}              
\newcommand{\vth}{\vartheta}            
\newcommand{\D}{{\rm d}}                
\newcommand{\II}{{\rm i}}               
\renewcommand{\Re}{{\rm Re\ }}          
\renewcommand{\Im}{{\rm Im\ }}          
\newcommand{\arcosh}{{\rm arcosh\,}}    
\newcommand{\erfc}{{\rm erfc\,}}        
\newcommand{\erfi}{{\rm erfi\,}}        
\newcommand{\erf}{{\rm erf\,}}          
\newcommand{\sign}{{\rm sign\,}}        
\newcommand{\sgn}{{\rm sgn\,}}          
\newcommand{\demi}{\frac{1}{2}}         
\newcommand{\wit}[1]{\widetilde{#1}}    
\newcommand{\wht}[1]{\widehat{#1}}      
\newcommand{\mel}[1]{\breve{#1}}        
\newcommand{\lap}[1]{\overline{#1}}     
\newcommand{\rar}{\rightarrow}          
\newcommand{\bra}[1]{\left\langle#1\right|}  
\newcommand{\ket}[1]{\left|#1\right\rangle}  
\newcommand{\fR}{\mathfrak{R}}          
\newcommand{\bR}{\lap{\mathfrak{R}}}    

\begin{titlepage} 
\begin{center}
{\LARGE\bf {Heterogeneous planar diffusion with axisymmetric power-law decaying diffusion coefficient under stochastic resetting}}
\end{center} 


\centering{\mbox{{\bf {Trifce Sandev}$^{a,b,c}$, {\bf S\'ebastien Fumeron}$^{d}$, {\bf Malte Henkel}$^{d,e}$ and {\bf Ervin K. Lenzi}$^f$}}}
\vskip 1.0 cm
\centerline{$^a${\it Research Center for Computer Science and Information Technologies,}} \centerline{\it Macedonian Academy of Sciences and Arts, Skopje, Macedonia}

\centerline{$^b$\textit{Institute of Physics, Faculty of Natural Sciences and Mathematics,}}  
\centerline{\textit{Ss. Cyril and Methodius University, Skopje, Macedonia}}

\centerline{$^c$\textit{Department of Physics, Korea University, Seoul, Republic of Korea}}

\centerline{$^d$\textit{Laboratoire de Physique et Chimie Th\'eoriques (CNRS UMR 7019), Universit\'e de Lorraine Nancy,}} 
\centerline{\textit{B.P. 70239,\: F - 54506 Vand{\oe}uvre-l\'es-Nancy Cedex, France}}

\centerline{$^e$\textit{Centro de F\'isica Te\'orica e Computacional, Universidade de Lisboa, Lisboa, Portugal}}

\centerline{$^f$\textit{Departamento de F\'isica, Universidade Estadual de Maring\'a, Maring\'a, PR, Brazil}}

\date{\today}
\begin{abstract}
We analyse two-dimensional isotropic heterogeneous diffusion with a
radially decaying diffusion coefficient $D(\rho)=D_0/\rho^2$, subject to
Poissonian stochastic resetting to the origin. Since multiplicative noise
makes the process interpretation-dependent, we treat the It\^o, Stratonovich, and H\"anggi--Klimontovich prescriptions within a single approach. For each, we obtain
the exact probability density function, the non-equilibrium stationary
state induced by resetting, the mean squared displacement, and
the first-passage properties. The non-equilibrium stationary
states are governed by modified Bessel functions and depart from the Laplace form found for one-dimensional homogeneous diffusion
under resetting, while the mean squared displacements saturate at long times as
$\langle\rho^2\rangle\sim\bar{r}^{-1/2}$, $\bar{r}$ being the resetting rate, contrary to the behaviour for the one-dimensional homogeneous diffusion, which is of form $\langle x^2\rangle\sim\bar{r}^{-1}$. The non-equilibrium stationary states also depart from those for the one-dimensional heterogeneous diffusion processes, while the corresponding mean squared displacements behave the same. The mean first-passage time to an absorbing boundary exhibits an optimal resetting rate at which it is
minimized. In addition, we incorporate memory effects through a subordination (continuous-time random walk) approach with a power-law kernel, which results from the long-tailed waiting time of the particle. Such a two-dimensional process with memory and resetting has not be analysed before in the literature. We derive the fractional Fokker--Planck equation together with the
non-equilibrium stationary
states, the mean squared displacements expressed via the three-parameter Mittag-Leffler function,
and the first-passage statistics. The heterogeneous diffusion process with memory in all three different interpretations of the multiplicative noise has not been analysed in the literature for the one-dimensional case, as well.

\end{abstract}

\end{titlepage}

\section{Introduction}

The diffusion in complex heterogeneous media often becomes anomalous due to different factors, such as memory effects, variation of the local diffusion coefficient, constrained motion in various geometries, power-law correlation in the driving noise, presence of velocity gradient, etc~\cite{report,review_chaos}. Anomalous diffusion is characterized by a power-law dependence of the mean squared displacement (MSD) on time, i.e., $\langle x^2(t)\rangle\sim t^\beta$, where $\beta\neq1$. The anomalous diffusion exponent $\beta$ defines sub-diffusion for $0<\beta<1$, normal diffusion for $\beta=1$ and super-diffusion for $\beta>1$~\cite{report,review_chaos}. 

The one-dimensional heterogeneous diffusion can be described by using the following Langevin equation~\cite{thesis,eli,review_chaos}
\begin{align}
    \dot{x}(t)=\sqrt{2\mathcal{D}(x)}\xi(t),
\end{align}
where $\mathcal{D}(x)$ is the position-dependent diffusion coefficient and $\xi(t)$ is white multiplicative noise with zero mean and $\delta$-correlation. Due to multiplicative noise, there are three distinct interpretations of heterogeneous diffusion that depend on how one integrates the Langevin equation. The corresponding Fokker-Planck equation for the probability density function (PDF) $P(x,t)$ becomes~\cite{thesis,eli}
\begin{align}
    \frac{\partial}{\partial t}P(x,t)=\frac{\partial}{\partial x}\left[\mathcal{D}^{\alpha}(x)\frac{\partial}{\partial x}\mathcal{D}^{1-\alpha}(x)P(x,t)\right],
\end{align}
with initial condition $P(x,0)=\delta(x-x_0)$. Here, for $\alpha=0$ one has the It\^{o} interpretation~\cite{ito}, for $\alpha=1/2$ it is Stratonovich interpretation~\cite{stratonovich} and for $\alpha=1$ it is the so-called H\"{a}nggi-Klimontovich interpretation~\cite{hkint,klim}. The one-dimensional case in all three interpretations has been widely investigated to describe heterogeneous diffusion in heterogeneous environment, crowded environment and random fractals~\cite{cherstvy,kaz,lov,proc,santos,srok}. However, higher-dimensional cases have rarely been  investigated~\cite{cherstvy2,ref1,ref2}. 

The isotropic two-dimensional heterogeneous diffusion can be described by the following Langevin equations~\cite{ref2}
\begin{align}\label{leq2}
    \dot{x}(t)&=\sqrt{2\mathcal{D}(x,y)}\,\xi_x(t),\nonumber\\ \dot{y}(t)&=\sqrt{2\mathcal{D}(x,y)}\,\xi_y(t),
\end{align}
where $\mathcal{D}(x,y)\equiv D(\rho)$ is the position-dependent diffusion coefficient, which, due to the isotropy, depends only on the radial coordinate $\rho=\sqrt{x^2+y^2}$ and is independent of the angular coordinate $\varphi$. This assumption applies to a wide range of problems, such as population ecology where animals disperse in a circular habitat around a central resource (central nest, watering hole, etc). Stochastic resetting captures homing, refuge-seeking or movement toward essential resources~\cite{Menon2025}. The Fokker-Planck equation for the radial PDF obtained from the Langevin equations~(\ref{leq2}), taking into consideration the radial symmetry, is derived in Ref.~\cite{ref2} and reads
\begin{align}
\frac{\partial}{\partial t}P(\rho,t)=\frac{1}{\rho}\frac{\partial}{\partial \rho}\left\{\rho \mathcal{D}^\alpha(\rho)\frac{\partial}{\partial \rho}\left[\mathcal{D}^{1-\alpha}(\rho)P(\rho,t)\right]\right\},
\end{align}
with initial condition $P(\rho,0)=\frac{1}{\rho}\delta(\rho)$. Here, three different interpretations of the heterogeneous diffusion will be analysed. Thus, throughout this work the stochastic-interpretation parameter is fixed by the convention
$\alpha=0$ (It\^{o}), $\alpha=1/2$ (Stratonovich), and $\alpha=1$ (H\"{a}nggi--Klimontovich). The corresponding radial Fokker-Planck equations are

\begin{align}
\frac{\partial}{\partial t}P(\rho,t)=\frac{1}{\rho}\frac{\partial}{\partial \rho}\left\{\rho\frac{\partial}{\partial \rho}\left[\mathcal{D}(\rho)P(\rho,t)\right]\right\}, \quad \alpha=0,
\end{align}

\begin{align}
\frac{\partial}{\partial t}P(\rho,t)=\frac{1}{\rho}\frac{\partial}{\partial \rho}\left\{\rho \sqrt{\mathcal{D}(\rho)}\frac{\partial}{\partial \rho}\left[\sqrt{\mathcal{D}(\rho)}P(\rho,t)\right]\right\}, \quad \alpha=1/2,
\end{align}

\begin{align}
\frac{\partial}{\partial t}P(\rho,t)=\frac{1}{\rho}\frac{\partial}{\partial \rho}\left\{\rho \mathcal{D}(\rho)\frac{\partial}{\partial \rho}P(\rho,t)\right\}, \quad \alpha=1.
\end{align}

In the last 15 years, after the seminal paper on diffusion under stochastic resetting by Evans and Majumdar~\cite{evans}, many different stochastic models have been generalized by considering resetting of the particle to its initial position, by which the systems reach non-equilibrium stationary states (NESSs)~\cite{evans,evans_majumdar_review,mendez,mendez2} while the mean first-passage times (MFPTs) become finite~\cite{evans,evans_majumdar_review,chapter,pal,shlomi,pal2,nayak}. The simplest case is the Poissonian resetting mechanism which means that the stochastic process is interrupted after a random time interval $\tau$ by resetting the particle to its initial
position, where the resetting probability is given by $p(\tau)=\bar{r}e^{-\bar{r}\tau}$ and $\bar{r}$ is the resetting rate. Therefore, such a stochastic process under Poissonian resetting can be described by the following renewal equation~\cite{mendez,evans_majumdar_review}
\begin{align}\label{renewal_eq}
    P_{\bar{r}}(x,t)=e^{-\bar{r}t}P(x,t)+\int_{0}^{t}\bar{r}e^{-\bar{r}t'}P(x,t')\,\D t',
\end{align}
where $P_{\bar{r}}(x,t)$ is the PDF in the presence of resetting, while $P(x,t)$ is the PDF of the corresponding process without resetting. 

Stochastic resetting has been investigated in many different contexts, such as in heterogeneous diffusion processes~\cite{lenzi_reset,hdp_reset,hdp2,hdp3}, anomalous diffusive~\cite{mendez,anomreset1,anomreset2} and ultra-slow diffusive processes~\cite{ultraslow}, rotational diffusion processes~\cite{irina}, reaction-diffusion processes \cite{durang}, telegrapher's processes~\cite{tgp1,tgp2}, run-and-tumble particle motion~\cite{tgp3,tgp4,tgp5}, geometric Brownian motion~\cite{gbm1,gbm2,gbm3}, random walks on networks~\cite{reset_networks,reset_networks2,reset_networks3,reset_networks4,reset_networks5}, quantum systems~\cite{quantum1,quantum2,quantum3}, etc. The experimental confirmation of the first passage
of Brownian motion under stochastic resetting, by using holographic
optical tweezers~\cite{exper1} and laser traps~\cite{exper2}, has also initiated new studies on various processes under stochastic resetting.

The paper is organized as follows. In Section~\ref{sec2}, we investigate the effects of resetting on the PDF, MSD and MFPT of the two-dimensional isotropic heterogeneous
diffusion process. We find that the system approaches a NESS in the long-time limit due to the resetting mechanism. We also observe an optimal resetting rate at which the MFPT is minimized. The corresponding heterogeneous diffusion process with memory is considered in Section~\ref{sec3} by introducing operational time. We analyze the effects of stochastic resetting on the corresponding PDF, NESS, and MSD. The first-passage properties are also investigated in detail. In Section~\ref{sec4} we summarize our findings.

\section{Effects of resetting: PDF, MSD and MFPT}\label{sec2}

In the present manuscript, we consider two-dimensional isotropic heterogeneous diffusion with a diffusion coefficient inversely proportional to the square of the radial coordinate, $D(\rho)=D_0/\rho^2$. Such models have been used to describe intracellular protein target-search, where macromolecular crowding yields a radially dependent diffusivity~\cite{cherstvy2,ref2}, with resetting mimicking protein dissociation/reassociation cycles~\cite{evans_majumdar_review}. Here, we note that in order to avoid the singularity at $\rho=0$ in the numerical simulations one uses the following behaviour of the diffusion coefficient $D(\rho)=D_0/(A+\rho^2)$, $A>0$~\cite{cherstvy,ref2,cherstvy2}.  

Under this assumption, the corresponding Fokker-Planck equation becomes~\cite{ref2}
\begin{align}\label{DgbmFP}
\frac{\partial}{\partial t}P(\rho,t)=\frac{D_0}{\rho}\frac{\partial}{\partial \rho}\left\{\rho^{1-2\alpha}\frac{\partial}{\partial \rho}\left[\rho^{-2(1-\alpha)}P(\rho,t)\right]\right\}
\end{align}
with initial condition $P(\rho,0)=\frac{1}{\rho}\delta(\rho)$. We use the angularly integrated radial convention in which the probability in the annulus $(\rho,\rho+d\rho)$ is $P(\rho,t)\,\rho\,d\rho$. Hence $\int_0^\infty P(\rho,t)\rho\,d\rho=1$, while the one-dimensional radial density with respect to $d\rho$ is $p_{\rm rad}(\rho,t)=\rho P(\rho,t)$. Therefore, we have the following Fokker-Planck equations for the three interpretations with $\alpha=\{0,1/2,1\}$, respectively,

\begin{align}\label{DgbmFP0}
\frac{\partial}{\partial t}P(\rho,t)=\frac{D_0}{\rho}\frac{\partial}{\partial \rho}\left\{\rho\frac{\partial}{\partial \rho}\left[\frac{1}{\rho^2}P(\rho,t)\right]\right\},
\end{align}

\begin{align}\label{DgbmFP05}
\frac{\partial}{\partial t}P(\rho,t)=\frac{D_0}{\rho}\frac{\partial}{\partial \rho}\left\{\frac{\partial}{\partial \rho}\left[\frac{1}{\rho}P(\rho,t)\right]\right\},
\end{align}

\begin{align}\label{DgbmFP1}
\frac{\partial}{\partial t}P(\rho,t)=\frac{D_0}{\rho}\frac{\partial}{\partial \rho}\left\{\frac{1}{\rho}\frac{\partial}{\partial \rho}P(\rho,t)\right\}.
\end{align}
Here we note that these three interpretations differ only by a drift term, since the It\^{o} and H\"anggi-Klimontovich interpretations can be rewritten in the Stratonovich form with drift terms. The It\^{o} form can be rewritten as

\begin{align}\label{DgbmFPitostr}
\frac{\partial}{\partial t}P(\rho,t)=\frac{D_0}{\rho}\frac{\partial}{\partial \rho}\left\{\frac{\partial}{\partial \rho}\left[\frac{1}{\rho}P(\rho,t)\right]\right\}-D_0\frac{\partial}{\partial\rho}\left[\frac{1}{\rho^2}P(\rho,t)\right],
\end{align} 
while H\"anggi-Klimontovich form as

\begin{align}\label{DgbmFPhkstr}
\frac{\partial}{\partial t}P(\rho,t)=\frac{D_0}{\rho}\frac{\partial}{\partial \rho}\left\{\frac{\partial}{\partial \rho}\left[\frac{1}{\rho}P(\rho,t)\right]\right\}+D_0\frac{\partial}{\partial\rho}\left[\frac{1}{\rho^2}P(\rho,t)\right].
\end{align} 

The solution of this equation is~\cite{ref2}
\begin{align}\label{PDF_no_reset}
    P(\rho,t)=\left\lbrace\begin{array}{lll}
    \frac{\rho^2}{4D_0t}\exp\left(-\frac{\rho^4}{16D_0t}\right),     & \alpha=0, \\ \\ 
    \frac{\rho}{2\Gamma(3/4)(D_0t)^{3/4}}\exp\left(-\frac{\rho^4}{16D_0t}\right),     & \alpha=1/2, \\ \\ 
    \frac{1}{\sqrt{\pi D_0t}}\exp\left(-\frac{\rho^4}{16D_0t}\right),    & \alpha=1,
    \end{array}\right.
\end{align}
with corresponding MSDs
\begin{align}\label{msd}    \langle\rho^2(t)\rangle=\sqrt{D_0t}\,\left\lbrace\begin{array}{lll}
        \sqrt{4\pi}, & \alpha=0, \\ \\
        \frac{\Gamma(1/4)}{\Gamma(3/4)}, & \alpha=1/2, \\ \\
        \frac{4}{\sqrt{\pi}}, & \alpha=1.
    \end{array}\right.
\end{align}

The singular point $\rho=0$ requires a boundary specification. For the backward generator
\[
\mathcal L_\alpha f(\rho)=\frac{D_0}{\rho^2}f''(\rho)+\frac{D_0(1-2\alpha)}{\rho^3}f'(\rho),
\]
the transformation $Y=\rho^4/(8D_0)$ yields the squared-Bessel process

$$
dY=(2-\alpha)\,dt+2\sqrt{Y}\,dW_t,$$
with dimension $\delta=2-\alpha$. Thus, for $\alpha=0$ ($\delta=2$) the origin is not reached from $\rho>0$ and acts as an entrance/polar boundary, whereas for $\alpha=\{1/2,1\}$ ($0<\delta<2$) it is a regular accessible boundary. In the free-propagator and resetting-to-origin sectors we use the standard non-absorbing squared-Bessel extension, for which trajectories can be initialized at the origin and leave it instantaneously. The first-passage problem introduced below avoids this singular boundary altogether by placing the absorbing target at a finite radius $a>0$.

Let us now consider the two-dimensional heterogeneous diffusion process governed by Eq.~(\ref{DgbmFP}) under Poissonian resetting. This means that the motion of a particle starting at $x=0$ at time $t=0$, which is
described by the PDF $P(\rho,t)$, is renewed at random times, following a distribution $p(\tau)=\bar{r}e^{-\bar{r}\tau}$, where $\bar{r}$ is the resetting rate. The dynamics of the particle under such Poissonian resetting is described by the renewal equation for the PDF $P_{\bar{r}}(\rho,t)$ in presence of resetting through the PDF $P(\rho,t)$ in
absence of resetting~\cite{mendez,evans_majumdar_review}
\begin{align}\label{renewal_eq}
    P_{\bar{r}}(\rho,t)=e^{-\bar{r}t}P(\rho,t)+\int_{0}^{t}\bar{r}e^{-\bar{r}t'}P(\rho,t')\,\D t'.
\end{align}
By Laplace transformation (the Laplace transform of the function $f(t)$ is defined by $\hat{f}(s)=\mathcal{L}[f(t)]=\int_{0}^{\infty}e^{-st}f(t)\,dt$), we have
 $$\hat{P}_{\bar{r}}(\rho,s)=\hat{P}(\rho,s+\bar{r})+\bar{r}s^{-1}\hat{P}(\rho,s+\bar{r}),$$
where we use the Laplace transformation of integral of a function, $\mathcal{L}\left[\int_{0}^{t}f(t')\,dt'\right]=s^{-1}\hat{f}(s)$ and the shift rule of the Laplace transform, $\mathcal{L}[e^{-at}f(t)]=\hat{f}(s+a)$. Therefore, in Laplace space the PDF reads
\begin{align}\label{connection}
    \hat{P}_{\bar{r}}(\rho,s)=\frac{s+\bar{r}}{s}\hat{P}(\rho,s+\bar{r}).
\end{align}
Let us derive the corresponding Fokker-Planck equation for such two-dimensional heterogeneous diffusion process under resetting. The Laplace transformation of eq.~(\ref{DgbmFP}) yields 
\begin{align}\label{DgbmFP_L}
s\hat{P}(\rho,s)-\frac{1}{\rho}\delta(\rho)=\frac{D_0}{\rho}\frac{\partial}{\partial \rho}\left\{\rho^{1-2\alpha}\frac{\partial}{\partial \rho}\left[\rho^{-2(1-\alpha)}\hat{P}(\rho,s)\right]\right\}.
\end{align}
By exchanging $s\rightarrow s+\bar{r}$, one finds
$$(s+\bar{r})\hat{P}(\rho,s+\bar{r})-\frac{1}{\rho}\delta(\rho)=\frac{D_0}{\rho}\frac{\partial}{\partial \rho}\left\{\rho^{1-2\alpha}\frac{\partial}{\partial \rho}\left[\rho^{-2(1-\alpha)}\hat{P}(\rho,s+\bar{r})\right]\right\}.$$We introduce a new function $\hat{P}_r(\rho,s)$ as defined by eq.~(\ref{connection}), from which we arrive at an equation of the form
\begin{align}\label{Dgbm_resetL}
s\hat{P}_{\bar{r}}(\rho,s)-\frac{1}{\rho}\delta(\rho)=\frac{D_0}{\rho}\frac{\partial}{\partial \rho}\left\{\rho^{1-2\alpha}\frac{\partial}{\partial \rho}\left[\rho^{-2(1-\alpha)}(\rho)P_{\bar{r}}(\rho,s)\right]\right\}-\bar{r}\hat{P}_{\bar{r}}(\rho,s)+s^{-1}\frac{\bar{r}}{\rho}\delta(\rho).
\end{align}
The inverse Laplace transform then yields
\begin{align}\label{Dgbm_reset}
\frac{\partial}{\partial t}P_{\bar{r}}(\rho,t)=\frac{D_0}{\rho}\frac{\partial}{\partial \rho}\left\{\rho^{1-2\alpha}\frac{\partial}{\partial \rho}\left[\rho^{-2(1-\alpha)}P_{\bar{r}}(\rho,t)\right]\right\}-\bar{r}P_{\bar{r}}(\rho,t)+\frac{\bar{r}}{\rho}\delta(\rho),
\end{align}
where the second term form the right-hand-side of the equation represents the loss of probability from the position $\rho$ due to the reset to the initial position $\rho_0=0$, while the third term is the gain of probability at $\rho_0=0$. We note that the solution of the Fokker-Planck equation~(\ref{Dgbm_reset}) is normalized, 
$$\int_{0}^{\infty}\hat{P}_{\bar{r}}(\rho,s)\rho\,d\rho=\frac{s+\bar{r}}{s}\int_{0}^{\infty}\tilde{P}(\rho,s+\bar{r})\rho\,d\rho=\frac{s+\bar{r}}{s}\frac{1}{s+\bar{r}}=\frac{1}{s},$$from where it follows that $\int_{0}^{\infty}P(\rho,t)\rho\,d\rho=1$. Here we use the normalization condition of the PDF in absence of resetting, i.e., $\int_{0}^{\infty}P(\rho,t)\rho\,d\rho=1$ and thus $\int_{0}^{\infty}\hat{P}(\rho,s)\rho\,d\rho=1/s$, which gives $\int_{0}^{\infty}\hat{P}(\rho,s+\bar{r})\rho\,d\rho=\frac{1}{s+\bar{r}}$. The normalization of the PDF $P_{\bar{r}}(\rho,t)$ can be shown by integration of the both sides of the Fokker-Planck equation~(\ref{Dgbm_reset}), i.e., $\int_{0}^{\infty}\cdots \rho\,d\rho$.

\subsection{PDF and NESS}
\label{PDF}

As we have already shown, the PDF in presence of resetting can be obtained from the renewal equation~(\ref{renewal_eq}) via the PDF~(\ref{PDF_no_reset}) in the absence of resetting. 

We list the three main cases. For $\alpha=0$, we find
\begin{align}\label{renewal_eq_0}
    P_{\bar{r}}(\rho,t)=\frac{\rho^2}{4D_0t}e^{-\left(\bar{r}+\frac{\rho^4}{16D_0t^2}\right)t}+\int_{0}^{t}\bar{r}\frac{\rho^2}{4D_0t}e^{-\left(\bar{r}+\frac{\rho^4}{16D_0t'^2}\right)t'}\,\D t'.
\end{align}
for $\alpha=1/2$, the PDF reads
\begin{align}\label{renewal_eq_0.5}
    P_{\bar{r}}(\rho,t)=\frac{\rho}{2\Gamma(3/4)(D_0t)^{3/4}}e^{-\left(\bar{r}+\frac{\rho^4}{16D_0t^2}\right)t}+\int_{0}^{t}\bar{r}\frac{\rho}{2\Gamma(3/4)(D_0t')^{3/4}}e^{-\left(\bar{r}+\frac{\rho^4}{16D_0t'^2}\right)t'}\,\D t',
\end{align}
while for $\alpha=1$, one has
\begin{align}\label{renewal_eq_1}
    P_{\bar{r}}(\rho,t)=\frac{1}{\sqrt{\pi D_0t}}e^{-\left(\bar{r}+\frac{\rho^4}{16D_0t^2}\right)t}+\int_{0}^{t}\bar{r}\frac{1}{\sqrt{\pi D_0t'}}e^{-\left(\bar{r}+\frac{\rho^4}{16D_0t'^2}\right)t'}\,\D t'.
\end{align}
These integrals have no closed solutions. In the long-time limit, one observes that the second terms in the renewal equations are dominant, while the first terms go to zero. Thus, for $t\rightarrow\infty$ the system approaches a non-equilibrium stationary state (NESS), which is given by
\begin{align}\label{ness_2}
    P^{st}_{\bar{r}}(x)=\lim_{t\rightarrow\infty}P_{\bar{r}}(\rho,t)=\lim_{s\rightarrow0}s\tilde{P}_{\bar{r}}(\rho,s)=\lim_{s\rightarrow0}s\frac{s+\bar{r}}{s}\tilde{P}(\rho,s+\bar{r})=\bar{r}\tilde{P}(\rho,\bar{r}).
\end{align}
Thus, for the three different interpretations, the NESSs have the form
\begin{align}\label{connection_ness}
    P^{st}_{\bar{r}}(x)=\left\lbrace\begin{array}{lll}
    \frac{\rho^2}{2}\left(\frac{\bar{r}}{D_0}\right)K_{0}\left(\frac{1}{2}\sqrt{\frac{\bar{r}}{D_0}}\rho^2\right),     & \alpha=0, \\ \\
    \frac{\rho^{3/2}}{\sqrt{2}\Gamma(3/4)}\left(\frac{\bar{r}}{D_0}\right)^{7/8}K_{1/4}\left(\frac{1}{2}\sqrt{\frac{\bar{r}}{D_0}}\rho^2\right),     & \alpha=1/2, \\ \\
    \left(\frac{\bar{r}}{D_0}\right)^{1/2}\exp\left(-\frac{1}{2}\sqrt{\frac{\bar{r}}{D_0}}\rho^2\right), & \alpha=1,
    \end{array}\right.
\end{align}
where $K_{\nu}(z)$ is the modified Bessel function of the second kind~\cite{abramowitz}. One can conclude that the NESSs differ from the NESS of the standard diffusion process under resetting, which is given by the Laplace distribution~\cite{evans}.

\begin{figure}
\centering
\includegraphics*[scale=.75,angle=0]{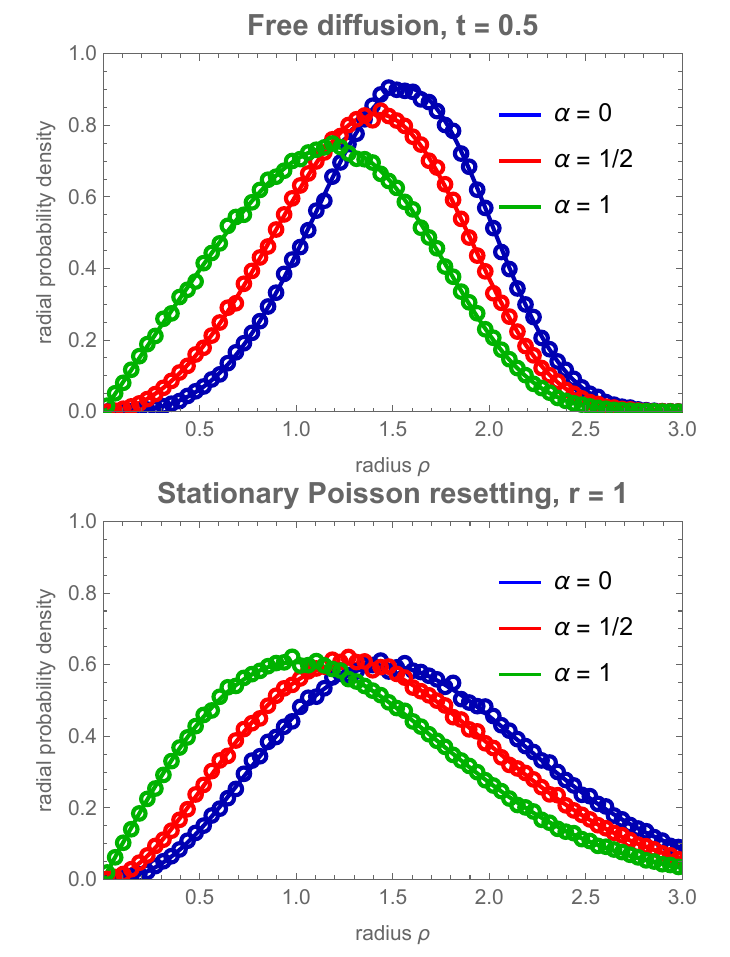}
\caption{Radial probability density $\rho P_{\bar{r}}(\rho,t)$ for
$\alpha=0$ (blue), $1/2$ (red), and $1$ (green).
Left panel: free propagators ($\bar{r}=0$) at $t=0.5$; Right panel: stationary densities
under resetting to the origin with $\bar{r}=1$.
We set $D_0=1$ and $N=250000$ samples were generated for all three interpretations of the multiplicative noise. Solid lines represent the analytical results, while open circles represent exact-transition Monte Carlo data (see Appendix~B.1 for details).}
\label{fig_pdf}
\end{figure}

The graphical representation of the radial PDF in absence and presence of resetting is given in Figure~\ref{fig_pdf}. The results are confirmed by Monte Carlo simulations. For each interpretation, $N=250000$ samples were generated
with pseudorandom seed 20261014.
The free distribution is sampled from a gamma law and the
stationary distribution from an exponentially mixed gamma law. Further details of the Monte Carlo sampling, numerical parameters,
and validation procedures are given in Appendix~B.

\subsection{MSD}

\begin{figure}[ht] 
\centering 
\includegraphics*[scale=.65,angle=0]{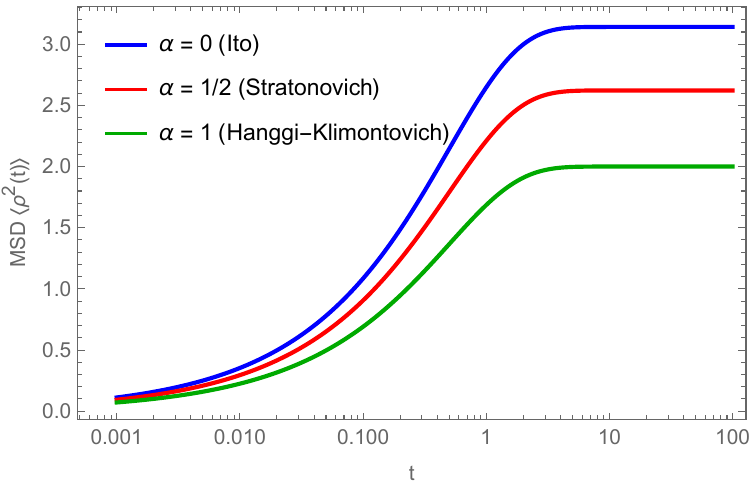} 
\caption{MSD~(\ref{msd_final_reset}) for $\alpha=0$ (blue line),  $\alpha=1/2$ (red line) and $\alpha=1$ (green line). We set $\bar{r}=1$ and $D_0=1$.} \label{fig_msd} 
\end{figure}

The MSD in the Laplace space is given by

\begin{align}
    \langle\hat{\rho}^2(s)\rangle_{\bar{r}}=\frac{s+\bar{r}}{s}\langle\hat{\rho}^2(s+\bar{r})\rangle_0=\frac{s+\bar{r}}{s}\sqrt{D_0}(s+\bar{r})^{-3/2}\,\left\lbrace\begin{array}{lll}
        \pi, & \alpha=0, \\ \\
        \frac{\sqrt{\pi}\Gamma(1/4)}{2\Gamma(3/4)}, & \alpha=1/2, \\ \\
        2, & \alpha=1,
    \end{array}\right.,
\end{align}
where for $\langle\hat{\rho}^2(s)\rangle_0$ we use the Laplace transformation of Eq.~(\ref{msd}). Thus, by inverse Laplace transformation, we obtain
\begin{align} \label{msd_final_reset}   \langle\rho^2(t)\rangle=\text{erf}\left(\sqrt{\bar{r}t}\right)\times\left\lbrace\begin{array}{lll}
        \pi\sqrt{\frac{D_0}{\bar{r}}}, & \alpha=0, \\ \\
        \pi\sqrt{\frac{D_0}{\bar{r}}}\frac{\Gamma(1/4)}{2\sqrt{\pi}\Gamma(3/4)}, & \alpha=1/2, \\ \\
        2\sqrt{\frac{D_0}{\bar{r}}}, & \alpha=1.
    \end{array}\right.,
\end{align}
In the limit $\bar{r}t\ll1$ the MSDs behave as in the case of no resetting, see eq.~(\ref{msd}), while in the limit $\bar{r}t\gg1$ they saturate and behave as $\langle\rho^2(t)\rangle\simeq r^{-1/2}$. Graphical representations of the MSDs~(\ref{msd_final_reset}) are given in Figure~\ref{fig_msd} from where one can observe saturation in the long time limit.

\subsection{First-passage properties}
\label{MFPT}

Here we will analyse the first-passage time of the process governed by Eq.~(\ref{Dgbm_reset}). The first-passage time is defined as the time required by the particle starting at $\rho = R$ to reach a target located at $\rho = a$, $a<R$ for the first time. The survival probability gives the probability that
the particle starting at $\rho=R$ has not reached the target located at $\rho = a$ up to time $t$. In order to find the survival probability in the presence of resetting $Q_{\bar{r}}(\rho,t)$, we should find the survival probability $Q_0(\rho,t)$ in the absence of resetting. We write the backward Fokker-Planck equation for the survival probability $Q_0(R,t)$, that is,

\begin{align}\label{DgbmFP_backward}
\frac{\partial}{\partial t}Q_0(\rho,t)=\frac{D_0}{\rho^{2(1-\alpha)+1}}\frac{\partial}{\partial \rho}\rho^{1-2\alpha}\frac{\partial}{\partial \rho}Q_0(\rho,t),
\end{align}
with initial condition $Q_{0}(\rho,0)=1$ and boundary conditions $Q_{0}(a,t)=0$ and $Q_{0}(\rho,t)\rightarrow1$ as $\rho\rightarrow\infty$.
The relation between the survival probabilities in the presence and absence of resetting, which is valid for a generic stochastic process, is given by~\cite{evans_majumdar_review,mendez,chapter}
\begin{align}\label{mfpt_surv_fpt}
    \hat{Q}_{\bar{r}}(R,s)=\frac{\hat{Q}_0(R, s+\bar{r})}{1-r\hat{Q}_0(R, s+\bar{r})}=\frac{1-\hat{F}_0(R,s+\bar{r})}{s+\bar{r}\hat{F}_0(R,s+\bar{r})},
\end{align}
where
\begin{align}
    F_0(R,t)=-\frac{\partial}{\partial t}Q_0(R,t) \quad \rightarrow \quad \hat{F}_0(R,s)=1-s\hat{Q}_0(R,s)
\end{align}
is the first-passage time density. The mean first-passage time, $\langle T_{\bar{r}}\rangle=-\int_{0}^{\infty}t\frac{d}{dt}Q_{\bar{r}}(R,t)\,dt=\hat{Q}(R,s=0)$, then can be calculated as~\cite{evans_majumdar_review,mendez,chapter}
\begin{align}
    \langle T_{\bar{r}}\rangle=\frac{\hat{Q}_{0}(R,\bar{r})}{1-\bar{r}\hat{Q}_0(R,\bar{r})}=\frac{1-\hat{F}_0(R,\bar{r})}{\bar{r}\hat{F}_0(R,\bar{r})}.
\end{align}

Let us solve Eq.~(\ref{DgbmFP_backward}). By Laplace transformation in respect to time $t$ and partial differentiation in respect to $\rho$, we have
\begin{align}
    \frac{\partial^2}{\partial\rho^2}\hat{Q}_{0}(\rho,s)+\frac{1-2\alpha}{\rho}\frac{\partial}{\partial\rho}\hat{Q}_{0}(\rho,s)-\frac{s}{D_0}\rho^2\hat{Q}_{0}(\rho,s)=-\frac{1}{D_0}\rho^2.
\end{align}
This homogeneous equation is a Bessel or Lommel-type equation with a solution

\begin{align}
    \hat{Q}_{0,h}(\rho,s)=C(s)\rho^{\alpha}K_{\alpha/2}\left(\frac{1}{2}\sqrt{\frac{s}{D_0}}\rho^2\right),
\end{align}
which decays as $\rho\rightarrow\infty$. The full solution must instead satisfy $\hat Q_0(\rho,s)\to1/s$ as $\rho\to\infty$.. The particular solution of the equation is $\hat{Q}_{0,p}(\rho,s)=1/s$. By taking into consideration the condition $\hat{Q}_{0}(a,s)=0$, we find the constant $C(s)$ form where we obtain the final solution

\begin{align}\label{survival_general}
    \hat{Q}_{0}(\rho,s)=\frac{1}{s}\left[1-\frac{\rho^{\alpha}K_{\alpha/2}\left(\frac{1}{2}\sqrt{\frac{s}{D_0}}\rho^2\right)}{a^{\alpha}K_{\alpha/2}\left(\frac{1}{2}\sqrt{\frac{s}{D_0}}a^2\right)}\right].
\end{align}
Therefore, for the MFPT, we find

\begin{align}
    \langle T_{\bar{r}}\rangle=\frac{1}{\bar{r}}\left[\frac{a^{\alpha}K_{\alpha/2}\left(\frac{1}{2}\sqrt{\frac{\bar{r}}{D_0}}a^2\right)}{R^{\alpha}K_{\alpha/2}\left(\frac{1}{2}\sqrt{\frac{\bar{r}}{D_0}}R^2\right)}-1\right].
\end{align}
Explicitly, we find the following. For $\alpha=0$, we obtain
\begin{align}\label{mfpt0}
    \langle T_{\bar{r}}\rangle=\frac{1}{\bar{r}}\left[\frac{K_{0}\left(\frac{1}{2}\sqrt{\frac{\bar{r}}{D_0}}a^2\right)}{K_{0}\left(\frac{1}{2}\sqrt{\frac{\bar{r}}{D_0}}R^2\right)}-1\right],
\end{align}
for $\alpha=1/2$ it follows
\begin{align}\label{mfpt05}
    \langle T_{\bar{r}}\rangle=\frac{1}{\bar{r}}\left[\frac{a^{1/2}K_{1/4}\left(\frac{1}{2}\sqrt{\frac{\bar{r}}{D_0}}a^2\right)}{R^{1/2}K_{1/4}\left(\frac{1}{2}\sqrt{\frac{\bar{r}}{D_0}}R^2\right)}-1\right],
\end{align}
while for $\alpha=1$, we find
\begin{align}\label{mfpt1}
    \langle T_{\bar{r}}\rangle=\frac{1}{\bar{r}}\left[\frac{a K_{1/2}\left(\frac{1}{2}\sqrt{\frac{\bar{r}}{D_0}}a^2\right)}{R K_{1/2}\left(\frac{1}{2}\sqrt{\frac{\bar{r}}{D_0}}R^2\right)}-1\right].
\end{align}

\begin{figure}[ht] 
\centering 
\includegraphics*[scale=.65,angle=0]{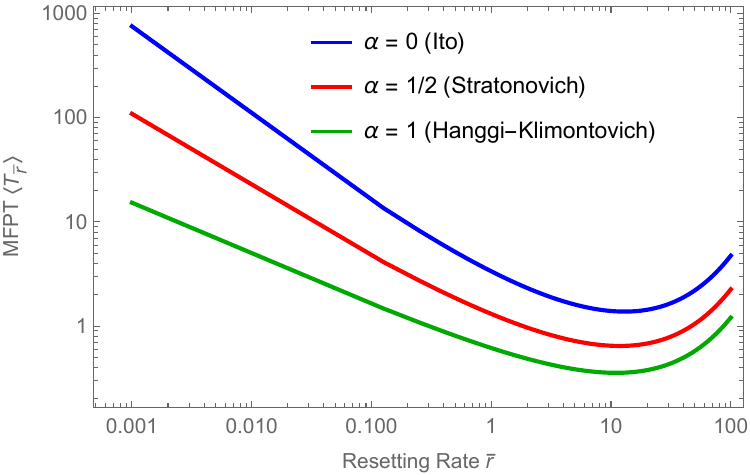} 
\caption{Mean first-passage time versus Poissonian resetting rate $\bar r$ for the absorbing disk of radius $a=0.2$ and resetting at $R=1$, with $D_0=1$. Curves correspond to $\alpha=0$ (blue), $\alpha=1/2$ (red), and $\alpha=1$ (green), calculated from Eqs.~\eqref{mfpt0},~\eqref{mfpt05}, and~\eqref{mfpt1}. The resetting rate interval extends to $\bar r=10^2$.} \label{fig1} 
\end{figure}

In Figure~\ref{fig1}, we give graphical representations of the dependence of MFPT on the resetting rate $\bar{r}$ for all three interpretations. We observe that there is an optimal resetting rate at which the MFPT is minimal. The optimal resetting rate at which the MFPT is minimal can be found from $\frac{\partial}{\partial\bar{r}}\langle T_{\bar{r}}\rangle=0$.
 
\subsection{Threshold time and loss of a finite threshold crossing}
\label{sec:thalf}

Beyond the mean first-passage time it is instructive to analyse, for the
resetting process of Section~\ref{PDF}, how long it takes for
half of the population to spread beyond the radius~$R$. This threshold
time~$t_{1/2}$ is defined by the condition
\begin{equation}
\Pi_{\bar r}(R,t_{1/2}) \;=\; \frac12,
\qquad
\Pi_{\bar r}(R,t) \;\equiv\; \int_0^R P_{\bar r}(\rho,t)\,\rho\,\mathrm{d}\rho ,
\label{eq:thalf-def}
\end{equation}
where $\Pi_{\bar r}(R,t)$ is the fraction of the (non-absorbed) population
found within radius $R$ at time $t$. We stress that $\Pi_{\bar r}$ is built from the free propagator under resetting of Section~\ref{PDF}, and is therefore distinct from the survival probability $Q_{\bar r}$ used for the MFPT in Section~\ref{MFPT}: here the boundary at $R$ is not absorbing and
probability may re-enter it.

For the reset-free process the cumulative mass admits the closed forms~\cite{ref2}
\begin{equation}
\Pi_0(R,t)=\int_0^R P(\rho,t)\,\rho\,\mathrm{d}\rho=
\begin{cases}
\displaystyle 1-\exp\!\left(-\dfrac{R^4}{16 D_0 t}\right), & \alpha=0,\\[10pt]
\displaystyle \frac{\gamma\!\left(\tfrac34,\ \tfrac{R^4}{16 D_0 t}\right)}{\Gamma(3/4)}, & \alpha=1/2,\\[10pt]
\displaystyle \operatorname{erf}\!\left(\dfrac{R^2}{4\sqrt{D_0 t}}\right), & \alpha=1,
\end{cases}
\label{eq:Pi0}
\end{equation}
where $\gamma(\alpha,z)$ is the lower incomplete gamma function. In the
presence of resetting, the renewal structure of Eq.~\eqref{renewal_eq} gives
$\Pi_{\bar r}(R,t)=e^{-\bar r t}\Pi_0(R,t)+\bar r\int_0^t
e^{-\bar r t'}\Pi_0(R,t')\,\mathrm{d}t'$, which, using
$e^{-\bar r t}+\bar r\int_0^t e^{-\bar r t'}\mathrm{d}t'=1$, reproduces
Eqs.~\eqref{renewal_eq_0}--\eqref{renewal_eq_1} and turns Eq.~\eqref{eq:thalf-def}
into a single transcendental equation for $t_{1/2}$.

In the vanishing-resetting limit, Eq.~\eqref{eq:thalf-def} reduces to
$\Pi_0(R,t_{1/2})=1/2$ and can be solved explicitly~\cite{ref2},
\begin{equation}
t_{1/2}^{(0)}=\frac{R^4}{16 D_0}\times
\begin{cases}
1/\ln 2\approx 1.4427, & \alpha=0,\\[4pt]
1/\theta^{\ast},\quad \theta^{\ast}\approx 0.4542, & \alpha=1/2,\\[4pt]
1/\big[\operatorname{erf}^{-1}(1/2)\big]^2\approx 4.3966, & \alpha=1,
\end{cases}
\label{eq:thalf-free}
\end{equation}
where $\theta^{\ast}$ is the median of the $\Gamma(3/4)$ distribution, i.e.
the root of $\gamma(3/4,\theta^{\ast})=\Gamma(3/4)/2$.

Switching on the resetting delays the escape monotonically: $t_{1/2}(\bar r)$
\emph{increases} with $\bar r$, in sharp contrast to the MFPT, which is
\emph{minimized} at a finite optimal rate. The reason is that resetting
continually returns probability to the origin, which lies inside $R$. As
$t\to\infty$ the fraction inside $R$ approaches the stationary value
\begin{equation}
\Pi_{\bar r}(R,\infty)=\bar r\,\widehat{\Pi}_0(R,\bar r)
\;\xrightarrow[\alpha=0]{}\;
1-z\,K_1(z),\qquad
z=2\sqrt{\tfrac{R^4}{16 D_0}\,\bar r}=\frac{R^2}{2}\sqrt{\frac{\bar r}{D_0}},
\label{eq:Pinf}
\end{equation}
with $K_\nu$ the modified Bessel function of the second kind and
$\widehat{\Pi}_0$ the Laplace transform of $\Pi_0$. A finite $t_{1/2}$ exists
only while $\Pi_{\bar r}(R,\infty)<1/2$. This is a threshold-existence criterion, not a thermodynamic phase transition: the qualitative change is simply the loss of a finite solution of Eq.~\eqref{eq:thalf-def}. The equality
$\Pi_{\bar r}(R,\infty)=1/2$ therefore defines a critical resetting
rate $\bar r^{\ast}$: for $\bar r>\bar r^{\ast}$ the stationary state retains
more than half of the population inside $R$ indefinitely, half of the
particles never cross the boundary, and $t_{1/2}$ ceases to exist. For the
It\^o case this criterion is fully explicit,
\begin{equation}
z^{\ast}K_1(z^{\ast})=\tfrac12
\;\Longrightarrow\;
z^{\ast}\approx 1.2572,
\qquad
\bar r^{\ast}=\frac{4 D_0}{R^4}\,(z^{\ast})^2 ,
\label{eq:rstar}
\end{equation}
giving, for $D_0=R=1$, $\bar r^{\ast}_{\alpha=0}\approx 6.32$, while a
numerical evaluation of Eq.~\eqref{eq:Pinf} yields
$\bar r^{\ast}_{\alpha=1/2}\approx 4.01$ and
$\bar r^{\ast}_{\alpha=1}\approx 1.92$. Since the approach to the stationary
state is exponential in the resetting clock, $t_{1/2}$ diverges
logarithmically as $\bar r\to\bar r^{\ast-}$.

\begin{figure}[t]
  \centering
  \includegraphics[width=0.7\linewidth]{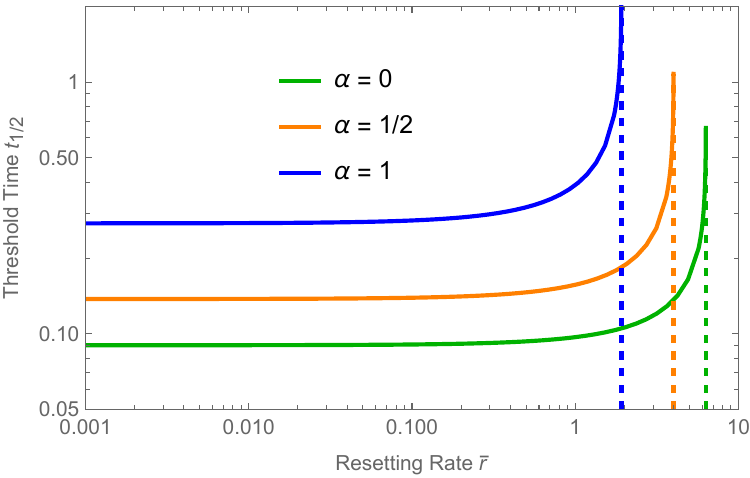}
  \caption{Threshold time $t_{1/2}$ versus resetting rate $\bar r$ for
    $\alpha=0$ (green), $\alpha=1/2$ (red) and $\alpha=1$ (blue), with
    $D_0=1$ and $R=1$. Each curve diverges at the critical resetting rate $\bar r^{\ast}$
    (dotted vertical lines): beyond $\bar r^{\ast}$ more than half of the
    population remains confined within $R$ and $t_{1/2}$ no longer exists.}
  \label{fig:thalf}
\end{figure}

Figure~\ref{fig:thalf} shows $t_{1/2}$ as a function of $\bar r$ for the three
interpretations. All curves depart from the reset-free values
\eqref{eq:thalf-free} at small $\bar r$, rise monotonically, and diverge at
their respective $\bar r^{\ast}$. The ordering
$\bar r^{\ast}_{\alpha=0}>\bar r^{\ast}_{\alpha=1/2}>\bar r^{\ast}_{\alpha=1}$
shows that the H\"anggi--Klimontovich prescription requires the weakest
resetting to trap half of the population. The threshold time and the MFPT
thus respond to resetting in opposite ways: the former is delayed without
bound and loses its finite threshold crossing at $\bar r^*$, while the latter is optimized at a finite rate, a distinction that has no counterpart in homogeneous
diffusion under resetting. 

\section{Memory effects and stochastic resetting}\label{sec3}

\subsection{Heterogeneous planar diffusion with memory}

We now generalise the model to the subordinated heterogeneous planar diffusion process by considering coupled Langevin equations by introducing operational time $u$~\cite{fogedby,baule}. For our model, we have the following coupled Langevin equations

\begin{align}
    \dot{\rho}(u)&=\sqrt{\frac{2D_0}{\rho^2}}\,\xi(u),\nonumber\\ \dot{t}(u)&=\zeta(u),
\end{align}
where $\xi(u)$ is the multiplicative white noise and $\zeta(u)$ is a completely one-sided L\'evy stable noise. The inverse process $\mathcal{S}(t)$ of the one-sided L\'evy stable process
$t(u)$ with the characteristic function $\left\langle e^{-st(u)}\right\rangle=e^{-\hat{\Psi}(s) u}$, where $\hat{\Psi}(s)$ is a ``characteristic exponent''. The corresponding Fokker-Planck equation for this subordinated process then reads

\begin{align}\label{DgbmFPmemory}
\frac{\partial}{\partial t}P_{\eta}(\rho,t)=\frac{d}{dt}\int_{0}^{t}\eta(t-t')\frac{D_0}{\rho}\frac{\partial}{\partial \rho}\left\{\rho^{1-2\alpha}\frac{\partial}{\partial \rho}\left[\rho^{-2(1-\alpha)}P_\eta(\rho,t')\right]\right\}\D t',
\end{align}
where $\eta(t)$ is a memory kernel such that $\eta(t)=\mathcal{L}^{-1}\left[1/\hat{\Psi}(s)\right]$. This process can be described in the framework of the continuous time random walk (CTRW) theory. The particle performing a heterogeneous planar diffusion with position-dependent diffusion coefficient can be trapped for some time at a given position due to, for example, crowded environment, before it continues to diffuse. By relating the particle's waiting time distribution, $\psi(t)$, to the memory kernel, $\eta(t)$, in Laplace space as $\hat{\psi}(s)=[1+1/\hat{\eta}(s)]^{-1} \sim 1-1/\hat{\eta}(s)$, one can derive the same generalized Fokker-Planck equation from CTRW theory. The random number of steps $n$ performed in such a CTRW process plays the role of the operational time~$u$~\cite{sokolov}.

Here, we will show that the solution of the generalized Fokker-Planck equation with memory kernel, which describes a non-Markovian process, can be found from the solution $P(\rho,u)$~(\ref{PDF_no_reset}) of eq.~(\ref{DgbmFP}). By Laplace transformation of eq.~(\ref{DgbmFP}), it follows

\begin{align}
s\hat{P}(\rho,s)-P(\rho,t=0)=\frac{D_0}{\rho}\frac{\partial}{\partial \rho}\left\{\rho^{1-2\alpha}\frac{\partial}{\partial \rho}\left[\rho^{-2(1-\alpha)}\hat{P}(\rho,s)\right]\right\}.
\end{align}
By exchanging $s\rightarrow 1/\hat{\eta}(s)$, one finds

\begin{align}\label{eqs}
\frac{1}{\hat{\eta}(s)}\hat{P}(\rho,1/\hat{\eta}(s))-P(\rho,t=0)=\frac{D_0}{\rho}\frac{\partial}{\partial \rho}\left\{\rho^{1-2\alpha}\frac{\partial}{\partial \rho}\left[\rho^{-2(1-\alpha)}\hat{P}(\rho,1/\hat{\eta}(s))\right]\right\}.
\end{align}
We introduce a new function $\hat{P}_\eta(\rho,s)$ defined by

\begin{align}\label{pdf subordination0}
     \hat{P}_\eta(\rho,s)  =\frac{1}{s \hat{\eta}(s)}\hat{P}\left(\rho,\frac{1}{\hat{\eta}(s)}\right)
\end{align}
in Eq.~(\ref{eqs}), from where we obtain

\begin{align}\label{DgbmFPmemoryLaplace}
s\hat{P}_{\eta}(\rho,s)-P(\rho,t=0)=s\hat{\eta}(s)\frac{D_0}{\rho}\frac{\partial}{\partial \rho}\left\{\rho^{1-2\alpha}\frac{\partial}{\partial \rho}\left[\rho^{-2(1-\alpha)}\hat{P}_\eta(\rho,s')\right]\right\}.
\end{align}
The inverse Laplace transformation yields the generalized Fokker-Planck equation~(\ref{DgbmFPmemory}). Thus, the PDF $P_{\eta}(\rho,t)$ can be obtained from the PDF $P(\rho,t)$. From Eq.~(\ref{pdf subordination0}), we have

\begin{align}\label{pdf subordination}
     \hat{P}_\eta(\rho,s)=\frac{1}{s \hat{\eta}(s)}\hat{P}\left(\rho,\frac{1}{\hat{\eta}(s)}\right)=\int_{0}^{\infty}P(\rho,u)\frac{1}{s \hat{\eta}(s)}e^{-\frac{u}{\hat{\eta}(s)}}\D u=\int_{0}^{\infty}P(\rho,u)\hat{h}(u,s)\,\D u,
\end{align}
where 

\begin{align}\label{sub_function}
     \hat{h}(u,s)=\frac{1}{s\hat{\eta}(s)}e^{-u/\hat{\eta}(s)}.
\end{align}
By inverse Laplace transformation, we find

\begin{align}\label{sub_integral}    P_\eta(\rho,t)=\int_{0}^{\infty}P(\rho,u)h(u,t)\,\D u.
\end{align}
This integral is known as decomposition integral. If the function $h(u,t)$ is non-negative, then it is called subordination function, while the integral~(\ref{sub_integral}) is known as subordination integral. This approach is known as subordination approach by which one can find the solution of a generalized Fokker-Planck equation with memory kernel by knowing the solution of the corresponding Fokker-Planck equation without memory kernel~\cite{report,barkai2,bazhlekova}. 

In order the subordination function $h(u,t)$ be a non-negative then its Laplace transform $\hat{h}(u,s)$ should be completely monotone function. It is a completely monotone if both, $\frac{1}{s\hat{\eta}(s)}$ and $e^{-u/\hat{\eta}(s)}$, are completely monotone. This means that $1/\hat{\eta}(s)$ should be a Bernstein function. Therefore, these conditions should satisfy the memory kernel for ensuring non-negativity of the solution $P_\eta(\rho,t)$ of the generalized Fokker-Planck equation. Moreover, it is normalized to 1 since 

$$\int_{0}^{\infty}\hat{P}_\eta(\rho,s)\rho\,d\rho=\frac{1}{s \hat{\eta}(s)}\int_{0}^{\infty}\hat{P}\left(\rho,\frac{1}{\hat{\eta}(s)}\right)\rho\,d\rho=\frac{1}{s \hat{\eta}(s)}\hat{\eta}(s)=\frac{1}{s},$$which yields $\int_{0}^{\infty}P_\eta(\rho,t)\rho\,d\rho=1$. Therefore, $P_\eta(\rho,t)$ is non-negative and normalized, i.e., it represents a PDF.

For a power-law memory kernel $\eta(t)=\frac{(t/\tau)^{\mu-1}}{\Gamma(\mu)}$, $0<\mu<1$, the generalized Fokker-Planck equation turns to the fractional Fokker-Planck equation

\begin{align}\label{fractionalDgbmFPmemory}
\frac{\partial}{\partial t}P_{\eta}(\rho,t)=\tau^{1-\mu}\,{_{RL}}D^{1-\mu}_{0+}\left(\frac{D_0}{\rho}\frac{\partial}{\partial \rho}\left\{\rho^{1-2\alpha}\frac{\partial}{\partial \rho}\left[\rho^{-2(1-\alpha)}P_\eta(\rho,t)\right]\right\}\right),
\end{align}
where 

$${_{RL}}D^{\kappa}_{0+}f(t)=\frac{1}{\Gamma(1-\kappa)}\frac{d}{dt}\int_{0}^{t}(t-t')^{-\kappa}f(t')\,dt'$$ is the Riemann-Liouville fractional derivative. We note that in the limit $\mu\rightarrow1$, we recover the standard heterogeneous diffusion equation~(\ref{DgbmFP}). 

For the power-law memory kernel, $\hat{\eta}(s)=\tau^{1-\mu}s^{-\mu}$, the noise $\zeta(u)$ is one-sided $\mu$-stable L\'evy noise with the stable index $0<\mu<1$, while $\mathcal{S}(t)$ is called the inverse-time $\mu$-stable subordinator~\cite{marcin}. The subordination function then becomes

\begin{align}\label{sub_function_levy}
    h(u,t)=\mathcal{L}^{-1}\left[\tau^{\mu-1}s^{\mu-1}e^{-u\tau^{\mu-1}s^{\mu}}\right]&=\frac{1}{\tau (t/\tau)^{\mu}}H_{1,1}^{1,0}\left[\frac{u/\tau}{(t/\tau)^{\mu}}\left|\begin{array}{c l}
    (1-\mu,\mu)\\
    (0,1)
\end{array}\right.\right]\nonumber\\&=\frac{t/\tau}{\mu u (u/\tau)^{1/\mu}}L_{\mu}\left(\frac{t/\tau}{(u/\tau)^{1/\mu}}\right),
\end{align}
where $L_{\mu}(z)$ is the one-sided L\'evy stable probability, which Laplace transform is $\mathcal{L}\left[L_{\mu}(t)\right]=e^{-s^{\mu}}$. Here, $H_{p,q}^{m,n}(z)$ is the Fox $H$-function defined by
means of the following Mellin-Barnes integral~\cite{book1,book2}

\begin{align}\label{H_integral}
H_{p,q}^{m,n}(z)=H_{p,q}^{m,n}\left[z\left|\begin{array}{c c}
    (a_1,A_1),...,(a_p,A_p)\\
    (b_1,B_1),...,(b_q,B_q)
  \end{array}\right.\right]=H_{p,q}^{m,n}\left[z\left|\begin{array}{c l}
    (a_p,A_p)\\
    (b_q,B_q)
  \end{array}\right.\right]=\frac{1}{2\pi\imath}\int_{\Omega}\theta(s)z^{-s}\,ds,
\end{align}
where

\begin{align}\label{theta_H}
    \theta(s)=\frac{\prod_{j=1}^{m}\Gamma(b_j+B_js)\prod_{j=1}^{n}
    \Gamma(1-a_j-A_js)}{\prod_{j=m+1}^{q}\Gamma(1-b_j-B_js)\prod_{j=n+1}^{p}\Gamma(a_j+A_js)},
\end{align}
$0\leq n\leq p$, $1\leq m\leq q$, $a_i,b_j \in \mathrm{C}$, $A_i,B_j\in\mathrm{R}^{+}$, $i=1,\dots,p$, $j=1,\dots,q$. Contour integration $\Omega$ starts at $c-\imath\infty$ and finishes at $c+\imath\infty$ separating the poles of the function $\Gamma(b_j+B_js)$, $j=1,\dots,m$ with those of the function $\Gamma(1-a_i-A_is)$, $i=1,\dots,n$. In the calculation of the subordination function~(\ref{sub_function_levy}), we use the relation between exponential function and Fox $H$-function,

\begin{align}\label{Hexp}
H_{0,1}^{1,0}\left[z\left|\begin{array}{cc}
-\\
(0,1)\end{array}\right.\right]=e^{-z},
\end{align}
as well as the Laplace transformation of the Fox $H$-function,

\begin{align}\label{Laplace H}
\mathcal{L}^{-1}\left[s^{-\rho}H_{p,q}^{m,n}\left[as^{\sigma}\left|\begin{array}{c
l}
    (a_p,A_p)\\
    (b_q,B_q)
  \end{array}\right.\right]\right]
  =t^{\rho-1}H_{p+1,q}^{m,n}\left[\frac{a}{t^{\sigma}}\left|\begin{array}{c
l}
(a_p,A_p),(\rho,\sigma)\\
    (b_q,B_q)
  \end{array}\right.\right].
\end{align}

From here we can also calculate the MSD of the subordinated process in Laplace space,

\begin{align}\label{mas subordination}
     \langle \hat{\rho}^2(s)\rangle_\eta=\frac{1}{s \hat{\eta}(s)}\left\langle \rho^2\left(\frac{1}{\hat{\eta}(s)}\right)\right\rangle_0=\sqrt{D_0}\Gamma(3/2)s^{-1}\sqrt{\hat{\eta}(s)}\,\left\lbrace\begin{array}{lll}
        \sqrt{4\pi}, & \alpha=0, \\ \\
        \frac{\Gamma(1/4)}{\Gamma(3/4)}, & \alpha=1/2, \\ \\
        \frac{4}{\sqrt{\pi}}, & \alpha=1.
    \end{array}\right.
\end{align}
where $\langle\hat{\rho}^2(s)\rangle_0$ is the MSD~(\ref{msd}) in the Laplace space.
For $\eta(t)=\frac{(t/\tau)^{\mu-1}}{\Gamma(\mu)}$, $0<\mu<1$, we obtain
\begin{align}\label{msd_mu}
     \langle \rho^2(t)\rangle_\eta&=C_\alpha\frac{\sqrt{\pi}}{2}\tau^{-\mu/2+1/2}\mathcal{L}^{-1}\left[s^{-\mu/2-1}\right]\nonumber\\&=\frac{\sqrt{\pi}}{2}\sqrt{D_0\tau}\frac{(t/\tau)^{\mu/2}}{\Gamma(1+\mu/2)}\left\lbrace\begin{array}{lll}
       \sqrt{4\pi},   & \alpha=0, \\ \\
       \Gamma(1/4)/\Gamma(3/4),   & \alpha=1/2, \\ \\
       4/\sqrt{\pi}, & \alpha=1.
     \end{array}\right.
\end{align}
For $\mu=1$, we recover the results for the MSD~(\ref{msd}) in the case of no memory in the system.

\subsection{Effects of stochastic resetting}

Resetting is imposed in physical time. At every Poissonian reset event the particle is returned to the reset point and the complete subordinated process is renewed, including the internal operational-time clock. This full-renewal prescription is essential because resetting and subordination do not generally commute.

In the presence of resetting, we use the renewal equation~\cite{mendez,evans_majumdar_review}

\begin{align}\label{renew_mu}
    P_{\eta,\bar{r}}(\rho,t)=e^{-\bar{r}t}P_{\eta}(\rho,t)+\bar{r}\int_{0}^{t}e^{-\bar{r}t'}P_{\eta}(\rho,t')\,\D t',
\end{align}
from where it follows

\begin{align}
    \hat{P}_{\eta,\bar{r}}(\rho,s)&=\frac{s+\bar{r}}{s}\hat{P}_{\eta}(\rho,s+\bar{r})=\frac{s+\bar{r}}{s}\frac{1}{(s+\bar{r})\hat{\eta}(s+\bar{r})}\hat{P}\left(\rho,\frac{1}{\hat{\eta}(s+\bar{r})}\right)\nonumber\\&=\frac{1}{s\hat{\eta}(s+\bar{r})}\hat{P}\left(\rho,\frac{1}{\hat{\eta}(s+\bar{r})}\right).
\end{align}
In the long-time limit one observes a NESS given by

\begin{align}
    P_{\eta,\bar{r}}^{st}(\rho)&=\lim_{s\rightarrow0}s\hat{P}_{\eta,\bar{r}}(\rho,s)=\frac{1}{\hat{\eta}(\bar{r})}\hat{P}\left(\rho,\frac{1}{\hat{\eta}(\bar{r})}\right).
\end{align}
For the power-law memory kernel $\eta(t)=\frac{(t/\tau)^{\mu-1}}{\Gamma(\mu)}$, $0<\mu<1$, the NESS becomes
\begin{align}
    P_{\eta,\bar{r}}^{st}(\rho)=\left\lbrace\begin{array}{lll}
    \frac{\rho^2}{2}\frac{(\bar{r}\tau)^{\mu}}{D_0\tau}K_{0}\left(\frac{1}{2}\sqrt{\frac{(\bar{r}\tau)^{\mu}}{D_0\tau}}\rho^2\right),     & \alpha=0, \\ \\
    \frac{\rho^{3/2}}{\sqrt{2}\Gamma(3/4)}\left(\frac{(\bar{r}\tau)^\mu}{D_0\tau}\right)^{7/8}K_{1/4}\left(\frac{1}{2}\sqrt{\frac{(\bar{r}\tau)^\mu}{D_0\tau}}\rho^2\right),     & \alpha=1/2, \\ \\
    \left(\frac{(\bar{r}\tau)^\mu}{D_0\tau}\right)^{1/2}\exp\left(-\frac{1}{2}\sqrt{\frac{(\bar{r}\tau)^\mu}{D_0\tau}}\rho^2\right), & \alpha=1.
    \end{array}\right.
\end{align}
From the renewal equation, we also obtain the MSD in the presence of memory and resetting. That is
\begin{align}
    \langle\hat{\rho}^2(s)\rangle_{\eta,r}=\frac{1}{s\hat{\eta}(s+\bar{r})}\langle\hat{\rho}^2(1/\hat{\eta}(s+\bar{r}))\rangle_{0}=\sqrt{D_0}s^{-1}\sqrt{\hat{\eta}(s+\bar{r})}\,\left\lbrace\begin{array}{lll}
        \pi, & \alpha=0, \\ \\
        \frac{\sqrt{\pi}\Gamma(1/4)}{2\Gamma(3/4)}, & \alpha=1/2, \\ \\
        2, & \alpha=1,
    \end{array}\right.
\end{align}
which in the long-time limit saturates to
\begin{align}
    \lim_{t\rightarrow\infty}\langle\rho^2(t)\rangle_{\eta,r}&=\lim_{s\rightarrow0}s\langle\hat{\rho}^2(s)\rangle_{\eta,r}\nonumber\\&=\frac{1}{\hat{\eta}(\bar{r})}\langle\hat{\rho}^2(1/\hat{\eta}(\bar{r}))\rangle_{0}=\sqrt{D_0}\sqrt{\hat{\eta}(\bar{r})}\,\left\lbrace\begin{array}{lll}
        \pi, & \alpha=0, \\ \\
        \frac{\sqrt{\pi}\Gamma(1/4)}{2\Gamma(3/4)}, & \alpha=1/2, \\ \\
        2, & \alpha=1.
    \end{array}\right.
\end{align}
For the power-law memory kernel, we obtain
\begin{align}\label{msd_mu_r}
     \langle \rho^2(t)\rangle_{\eta,\bar{r}}&=C_\alpha\frac{\sqrt{\pi}}{2}\tau^{-\mu/2+1/2}\mathcal{L}^{-1}\left[\frac{s^{-1}}{(s+\bar{r})^{\mu/2}}\right]\nonumber\\&=\frac{\sqrt{\pi}}{2}\sqrt{D_0\tau}(t/\tau)^{\mu/2}E_{1,\mu/2+1}^{\mu/2}(-\bar{r}t)\left\lbrace\begin{array}{lll}
       \sqrt{4\pi},   & \alpha=0, \\ \\
       \Gamma(1/4)/\Gamma(3/4),   & \alpha=1/2, \\ \\
       4/\sqrt{\pi}, & \alpha=1,
     \end{array}\right.
\end{align}
where $E_{\alpha,\beta}^{\gamma}(z)$ is the three-parameter Mittag-Leffler function~\cite{prabhakar}

\begin{equation}\label{three parameter ML}
E_{\alpha,\beta}^\gamma(z)=\sum_{k=0}^{\infty}\frac{(\gamma)_k}{\Gamma(\alpha k+\beta)}\frac{z^k}{k!},
\end{equation}
where $\beta, \gamma, z \in \mathbb{C}$, $\Re(\alpha)>0$, $(\gamma)_{k}$ is the Pochhammer symbol. Its Laplace transform reads~\cite{prabhakar}
\begin{equation}\label{Laplace ML3_1}
\mathcal{L}\left[t^{\beta-1}E_{\alpha, \beta}^{\gamma}\left(-\lambda t^{\alpha}\right)\right](s)
=\frac{s^{\alpha\gamma-\beta}}{(s^\alpha+\lambda)^\gamma},
\end{equation}
where $|\lambda/s^{\alpha}|<1$.
For the long-time limit, the MSD saturates and behaves as $\langle \rho^2(t)\rangle_{\eta,\bar{r}}\sim \bar{r}^{-\mu/2}$, i.e., 

\begin{align}\label{msd_mu_r_asympt}
     \langle \rho^2(t)\rangle_{\eta,\bar{r}}=\sqrt{\frac{D_0\tau}{(\bar{r}\tau)^{\mu}}}\left\lbrace\begin{array}{lll}
       \pi,   & \alpha=0, \\ \\
       \frac{\sqrt{\pi}\Gamma(1/4)}{2\Gamma(3/4)},   & \alpha=1/2, \\ \\
       2, & \alpha=1,
     \end{array}\right.
\end{align}
Here, we use the asymptotic behaviour of the three-parameter Mittag-Leffler function \cite{garra},

\begin{equation}\label{GML_formula3}
E_{\alpha,\beta}^{\gamma}(-z)=\frac{z^{-\gamma}}{\Gamma(\gamma)}\sum_{n=0}^{\infty}\frac{\Gamma(\gamma+n)}{\Gamma(\beta-\alpha(\gamma+n))}\frac{(-z)^{-n}}{n!}, \quad z>1,
\end{equation}
where $\alpha<2$.

\begin{figure}[ht] 
\centering 
\includegraphics*[scale=.65,angle=0]{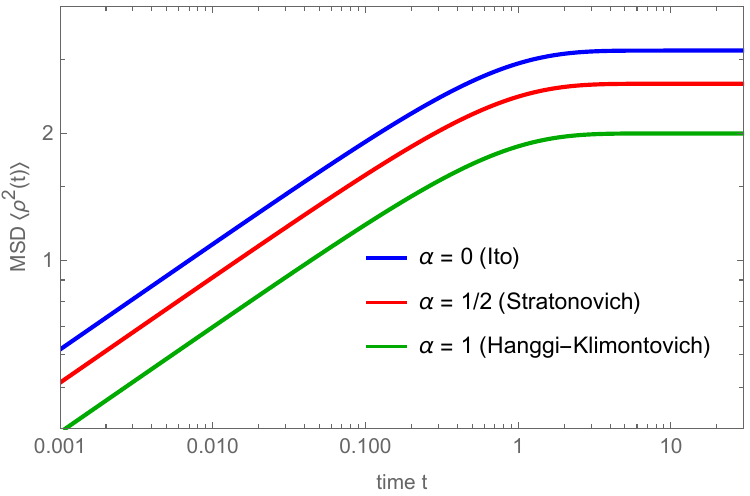} 
\caption{MSD~(\ref{msd_mu_r}) for $\alpha=0$ (blue line),  $\alpha=1/2$ (red line) and $\alpha=1$ (green line). We set $\bar{r}=1$, $D_0=1$ and $\mu=1/2$.} \label{fig_msd_mu} 
\end{figure}

A graphical representation of the MSD~(\ref{msd_mu_r}) is given in Figure~\ref{fig_msd_mu}, which shows the expected early-time anomalous growth \(\langle\rho^2(t)\rangle\sim t^{\mu/2}\) for early times to saturation for late times, as it was expected due to the resetting mechanism.

\subsection{First-passage properties}

In order to find the MFPT in presence of resetting we shall first find the survival probability in absence of resetting for the corresponding process with memory. Here we will find the survival probability that gives the probability that
the particle starting at $\rho=R$ has not reached the target located at $\rho = a$ up to time $t$. Thus, we analyse the backward Fokker-Planck equation from the survival probability

\begin{align}\label{DgbmFP_backward_memory}
\frac{\partial}{\partial t}Q_\eta(\rho,t)=\frac{d}{dt}\int_{0}^{t}\eta(t-t')\frac{D_0}{\rho^{2(1-\alpha)+1}}\frac{\partial}{\partial \rho}\rho^{1-2\alpha}\frac{\partial}{\partial \rho}Q_\eta(\rho,t')\,dt',
\end{align}
with initial condition $Q_{\eta}(\rho,0)=1$ and boundary conditions $Q_{\eta}(a,t)=0$ and $Q_{\eta}(\rho,t)\rightarrow1$ as $\rho\rightarrow\infty$. By using the subordination approach, in the same way as it was done for the PDF in Section~\ref{sec3}, we find

\begin{align}\label{sub_integral_surv}    Q_\eta(\rho,t)=\int_{0}^{\infty}Q_0(\rho,u)h(u,t)\,\D u.
\end{align}
Here $h(u,t)$ is the subordination function which in Laplace space is given by Eq.~(\ref{sub_function}). In Laplace space, one finds

\begin{align}\label{surv}
     \hat{Q}_\eta(\rho,s)&=\int_{0}^{\infty}Q_0(\rho,u)\hat{h}(u,s)\,\D u=\frac{1}{s \hat{\eta}(s)}\hat{Q}_0\left(\rho,\frac{1}{\hat{\eta}(s)}\right)\nonumber\\&=\frac{1}{s}\left[1-\frac{\rho^{\alpha}K_{\alpha/2}\left(\frac{1}{2}\sqrt{\frac{1}{D_0\hat{\eta}(s)}}\rho^2\right)}{a^{\alpha}K_{\alpha/2}\left(\frac{1}{2}\sqrt{\frac{1}{D_0\hat{\eta}(s)}}a^2\right)}\right].
\end{align}
Therefore, the MFPT becomes

\begin{align}\label{mfpt_general_eta}
    \langle T_{\eta,\bar{r}}\rangle=\frac{\hat{Q}_{\eta}(R,\bar{r})}{1-\bar{r}\hat{Q}_{\eta}(R,\bar{r})}=\frac{1}{\bar{r}}\left[\frac{a^{\alpha}K_{\alpha/2}\left(\frac{1}{2}\sqrt{\frac{1}{D_0\hat{\eta}(\bar{r})}}a^2\right)}{R^{\alpha}K_{\alpha/2}\left(\frac{1}{2}\sqrt{\frac{1}{D_0\hat{\eta}(\bar{r})}}R^2\right)}-1\right].
\end{align}

\begin{figure}[ht] 
\centering 
\includegraphics*[scale=.65,angle=0]{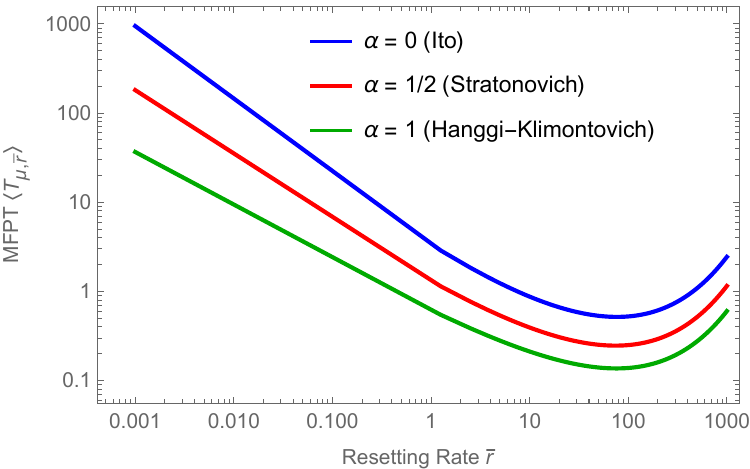} 
\caption{Mean first-passage time versus resetting rate $\bar r$ for $\mu=3/4$, $D_0=1$, $R=1$, $\tau=1$, and absorbing-disk radius $a=0.2$. The full process is reset at $R$ in physical time. The blue, red and green curves correspond to $\alpha=0$, $1/2$ and $1$, respectively; see Eqs.~\eqref{mfpt0mu}, ~\eqref{mfpt05mu}, and~\eqref{mfpt1mu}. The resetting rate interval extends to $\bar r=10^3$.} \label{fig2} 
\end{figure}

\begin{figure}[ht] 
\centering 
\includegraphics*[scale=.65,angle=0]{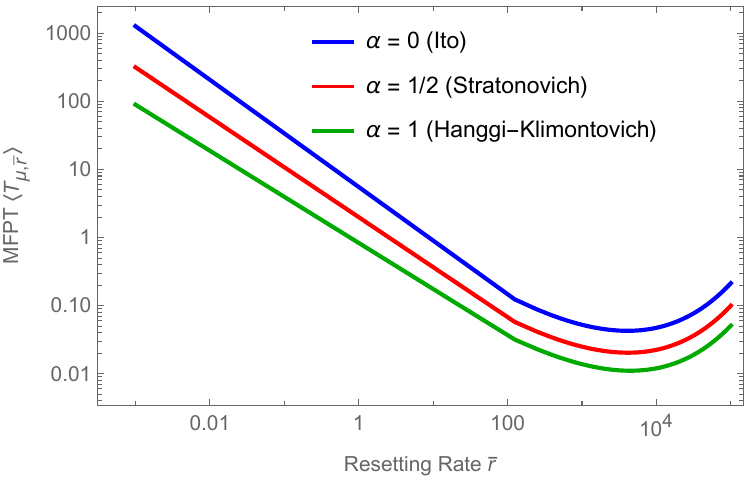} 
\caption{Mean first-passage time versus resetting rate $\bar r$ for $\mu=1/2$, $D_0=1$, R=1$, \tau=1$, and absorbing-disk radius $a=0.2$. The full process is reset at $R$ in physical time. The blue, red and green curves correspond to $\alpha=0$, $1/2$ and $1$, respectively; see Eqs.~\eqref{mfpt0mu}, ~\eqref{mfpt05mu}, and~\eqref{mfpt1mu}. The resetting rate interval extends to $\bar r=10^5$.} \label{fig3} 
\end{figure}

For the power-law memory kernel $\eta(t)=\frac{(t/\tau)^{\mu-1}}{\Gamma(\mu)}$, $0<\mu<1$, for MFPT $\langle T_{\eta,\bar{r}}\rangle\rightarrow\langle T_{\mu,\bar{r}}\rangle$, we obtain

\begin{align}\label{mfpt0mu}
    \langle T_{\mu,\bar{r}}\rangle=\frac{1}{\bar{r}}\left[\frac{K_{0}\left(\frac{1}{2}\sqrt{\frac{(\bar{r}\tau)^\mu}{D_0\tau}}a^2\right)}{K_{0}\left(\frac{1}{2}\sqrt{\frac{(\bar{r}\tau)^\mu}{D_0\tau}}R^2\right)}-1\right]
\end{align}
for $\alpha=0$,

\begin{align}\label{mfpt05mu}
    \langle T_{\mu,\bar{r}}\rangle=\frac{1}{\bar{r}}\left[\frac{a^{1/2}K_{1/4}\left(\frac{1}{2}\sqrt{\frac{(\bar{r}\tau)^\mu}{D_0\tau}}a^2\right)}{R^{1/2}K_{1/4}\left(\frac{1}{2}\sqrt{\frac{(\bar{r}\tau)^\mu}{D_0\tau}}R^2\right)}-1\right]
\end{align}
for $\alpha=1/2$, and

\begin{align}\label{mfpt1mu}
    \langle T_{\mu,\bar{r}}\rangle=\frac{1}{\bar{r}}\left[\frac{aK_{1/2}\left(\frac{1}{2}\sqrt{\frac{(\bar{r}\tau)^\mu}{D_0\tau}}a^2\right)}{RK_{1/2}\left(\frac{1}{2}\sqrt{\frac{(\bar{r}\tau)^\mu}{D_0\tau}}R^2\right)}-1\right]
\end{align}
for $\alpha=1$. The Markovian limit follows immediately: as $\mu\to1$, $1/\hat\eta(s)=(s\tau)^\mu/\tau\to s$, so Eqs.~\eqref{mfpt0mu}--\eqref{mfpt1mu} reduce exactly to Eqs.~\eqref{mfpt0}--\eqref{mfpt1}. In Figures~\ref{fig2} and~\ref{fig3}, we give graphical representations of the dependence of MFPT on the resetting rate $\bar{r}$ for all three interpretations for different values of $\mu$. One observes that there is an optimal resetting rate at which the MFPT is minimal. If we compare all three Figures~\ref{fig1},~\ref{fig2} and~\ref{fig3}, we see that for relatively close target, by decreasing parameter $\mu$ the optimal resetting rate increases. This is an expected result since as $\mu$ decreases, the waiting time of the particle increases, so more frequent resets are needed in order to avoid particle trapping before it reaches the target. The numerical procedures used to determine the optimal resetting rates,
the minimum MFPTs, and the threshold times are described in Appendix~B.

\subsection{Threshold time and loss of a finite threshold crossing}
\label{sec:thalfmu}

Here we can also analyse how long it takes for
half of the population to spread beyond the radius~$R$, such that the threshold
time~$t_{1/2}$ can be found from the condition
\begin{equation}
\Pi_{\eta,\bar r}(R,t_{1/2}) \;=\; \frac12,
\qquad
\Pi_{\eta,\bar r}(R,t) \;\equiv\; \int_0^R P_{\eta,\bar r}(\rho,t)\,\rho\,\mathrm{d}\rho .
\label{eq:thalf-defmu}
\end{equation}
Due to the renewal equation~(\ref{renew_mu}), we have 

\begin{align}\label{renew_pi_mu}
    \Pi_{\eta,\bar r}(R,t)=e^{-\bar r t}\Pi_{\eta}(R,t)+\bar r\int_0^t
e^{-\bar r t'}\Pi_{\eta}(R,t')\,\mathrm{d}t',
\end{align}
where 

\begin{align}
    \Pi_{\eta}(R,t) \;\equiv\; \int_0^R P_{\eta}(\rho,t)\,\rho\,\mathrm{d}\rho.
\end{align}
In Laplace space it reads

\begin{align}
    \hat{\Pi}_{\eta}(R,s) \;\equiv\; \int_0^R \hat{P}_{\eta}(\rho,s)\,\rho\,\mathrm{d}\rho=\frac{1}{s\hat{\eta}(s)}\int_{0}^{R}\hat{P}(\rho,1/\hat{\eta}(s))\rho d\rho.
\end{align}

For a power-law memory kernel $\eta(t)=(t/\tau)^{\mu-1}/\Gamma(\mu)$, $0<\mu<1$, from eq.~(\ref{PDF_no_reset}), for $\alpha=0$, we obtain

\begin{align}
    \Pi_{\eta}(R,t) =1-H_{1,2}^{2,0}\left[\left.\frac{R^4}{16D_0\tau(t/\tau)^\mu}\right|\begin{array}{cc}
    (1,\mu)  \\
    (1,1), (0,1) 
    \end{array}\right],
\end{align}    
where $H_{p,q}^{m,n}(z)$ is the Fox $H$-function~(\ref{H_integral}). For $\alpha=1/2$, we find

\begin{align}
    \Pi_{\eta}(R,t) =1-\frac{1}{\Gamma(3/4)}H_{1,2}^{2,0}\left[\left.\frac{R^4}{16D_0\tau(t/\tau)^\mu}\right|\begin{array}{cc}
    (1,\mu)  \\
    (3/4,1), (0,1) 
    \end{array}\right],
\end{align}    
and for $\alpha=1$, 

\begin{align}
    \Pi_{\eta}(R,t) =1-H_{1,1}^{1,0}\left[\left.\frac{R^2}{\sqrt{16D_0\tau(t/\tau)^\mu}}\right|\begin{array}{cc}
    (1,\mu/2)  \\
    (0,1) 
    \end{array}\right].
\end{align} 
From these exact results for $\Pi_{\eta}(R,t)$ and the renewal equation~(\ref{renew_pi_mu}) for $\Pi_{\eta,\bar{r}}(R,t)$ one can numerically estimate $t_{1/2}$.

\section{Summary}\label{sec4}

In this work we present detailed analytical results on the stochastic resetting problem of two dimensional isotropic planar diffusion in heterogeneous media, represented by axisymmetric power-law decaying diffusivity. We consider three different interpretations of the multiplicative noise, It\^{o}, Stratonovich and H\"{a}nggi-Klimontovich. We obtained exact results for the NESS reached in the long time limit due to the resetting, and investigate the effects of resetting on the MSD. We also analysed in detail the first-passage properties by solving the backward Fokker-Planck equation for the survival probability. We calculated the survival probability, the first-passage density and the mean first passage time and showed that there is an optimal resetting rate, which makes the MFPT minimal. The threshold time, which defines how long it takes for half of the population to spread beyond some radius of interest in the plane, was also investigated. The memory effects of the environment are also investigated in the framework of the subordination approach by introdusing . The interplay between heterogeneity, memory, and resetting are analysed, as well. Such a heterogeneous diffusion model with memory in all three different interpretations has not been investigated neither for one-dimensional case nor for higher dimensional case. The present model could be of interest for a description of two-dimensional diffusion in heterogeneous media and disordered structures.      

We also distinguish ensemble observables such as the MSD from time-averaged observables. Resetting can generate stationary ensemble statistics while the time-averaged MSD and its fluctuations retain a dependence on the observation protocol and on whether the process is initialized in a fresh or equilibrated resetting ensemble; see, e.g., Ref.~\cite{gbm2}. A complete TAMSD and ergodicity-breaking analysis for the present two-dimensional heterogeneous model is beyond the scope of the present work and is left for future study.

The investigation of the two- and three-dimensional heterogeneous diffusion processes with general position-dependent diffusion coefficient $D(r)\sim r^\alpha$, $\alpha<2$, needs special attention and we leave for future research. Investigation of the (non)ergodic properties of such two-dimensional heterogeneous diffusion processes will be of interest for further research, generalizing the well-known result for the one-dimensional heterogeneous diffusion processes~\cite{cherstvy,cherstvy2,hdp2,ergod1}.


\appendix

\section{Backward first-passage equation and Bessel solution}

For the resetting protocol used in Sections~\ref{MFPT} and 3.3, let $u_s(\rho;a)=\mathbb E_\rho[e^{-sT_a}]$, where $T_a$ is the first hitting time of the absorbing radius $a$. In the interval $\rho>a$, $u_s$ satisfies

\begin{equation}
\frac{D_0}{\rho^2}u_s''(\rho)+\frac{D_0(1-2\alpha)}{\rho^3}u_s'(\rho)=s u_s(\rho),
\qquad u_s(a;a)=1,\qquad u_s(\rho;a)\to0\quad(\rho\to\infty).
\end{equation}
With $z=(\rho^2/2)\sqrt{s/D_0}$ and $u_s(\rho;a)=\rho^\alpha v(z)$, the equation reduces to the modified Bessel equation

\begin{equation}
z^2v''+zv'-(z^2+\alpha^2/4)v=0.
\end{equation}
Hence $v=A I_{\alpha/2}(z)+B K_{\alpha/2}(z)$. The $I$ branch is excluded (A=0) because it grows exponentially for $z\to\infty$, while the $K$ branch decays. Imposing $u_s(a;a)=1$ gives

\begin{equation}
u_s(R;a)=\frac{R^\alpha K_{\alpha/2}\!\left(\frac{1}{2}\sqrt{\frac{s}{D_0}}R^2\right)}{a^\alpha K_{\alpha/2}\!\left(\frac{1}{2}\sqrt{\frac{s}{D_0}}a^2\right)}.
\end{equation}
For complete Poissonian resetting to $R$ at rate $\bar r$, the standard renewal identity then yields

\begin{equation}
\langle T_{\bar r}\rangle=\frac{1-u_{\bar r}(R;a)}{\bar r\,u_{\bar r}(R;a)}.
\end{equation}
For the power-law memory kernel, subordination replaces $s$ by $\Psi(s)=1/\hat\eta(s)=(s\tau)^\mu/\tau$. With full renewal in physical time, this gives

\begin{equation}
\langle T_{\mu,\bar r}\rangle=\frac{1-u_{\Psi(\bar r)}(R;a)}{\bar r\,u_{\Psi(\bar r)}(R;a)},
\end{equation}
which reduces to the Markovian expression when $\mu\to1$.

\section{Numerical implementation and validation}
\label{app:numerical}

The analytical results reported in the main text were supplemented by numerical
calculations designed to test the probability densities, the threshold-time
construction, and the first-passage expressions. Unless otherwise stated, the
dimensionless parameter choice used in the figures is $D_0=1$. For the
first-passage calculations we further set $R=1$, $a=0.2$, and, in the
non-Markovian case, $\tau=1$. All radial probability densities are represented
with respect to $d\rho$, i.e., as $\rho P(\rho,t)$.

\subsection{Exact-transition Monte Carlo validation of the radial densities}

For the free process, the transformation

\begin{equation}
Y=\frac{\rho^4}{16D_0 t}
\end{equation}
shows that

\begin{equation}
Y\sim {\rm Gamma}(k_\alpha,1),
\qquad
k_\alpha=1-\frac{\alpha}{2}.
\end{equation}
Therefore, an exact sample of the radial position at a fixed time $t$ can be
generated as

\begin{equation}
\rho(t)=\left(16D_0 t\,G_\alpha\right)^{1/4},
\qquad
G_\alpha\sim {\rm Gamma}(k_\alpha,1).
\label{eq:MC_free}
\end{equation}
For the stationary resetting process, the age $U$ since the most recent reset is
exponentially distributed,

\begin{equation}
U\sim {\rm Exp}(\bar r),
\end{equation}
and is independent of $G_\alpha$. Hence the stationary radial position can be
sampled exactly from

\begin{equation}
\rho_{\rm st}
=
\left(16D_0 U G_\alpha\right)^{1/4}.
\label{eq:MC_stationary}
\end{equation}
Equations~\eqref{eq:MC_free} and \eqref{eq:MC_stationary} provide an
exact-transition Monte Carlo implementation of the stochastic process and do not
require a time discretization through the singular point $\rho=0$.

For Figure~1, $N=250000$ independent samples were generated for each
interpretation using pseudorandom seed 20261014, with $D_0=1$, $t=0.5$ for the
free process, and $\bar r=1$ for the stationary process. Histograms were
constructed with 84 equally spaced bins on $0\leq\rho\leq3.5$. The analytical
curves were evaluated on 450 radial points over
$0.002\leq\rho\leq3.5$. As an additional check, empirical and analytical
cumulative probabilities were compared at
$R=0.5,1,1.5,$ and $2$. Over all three stochastic interpretations and both the
free and stationary cases, the largest absolute discrepancy was
$2.41\times10^{-3}$.

Because the probability-density validation is based on exact transition laws,
the regularization $D(\rho)=D_0/(A+\rho^2)$ is not required for Figure~1.
Such a regularization is useful for direct time-stepping schemes near the
singular origin, but the numerical data reported in Figure~1 do not depend on a
regularization parameter $A$.

\subsection{Evaluation of the MSD}

The Markovian MSD in Figure~2 was evaluated directly from the closed expression
derived in the main text. The time interval
$10^{-3}\leq t\leq10^2$ was sampled with 500 logarithmically spaced points,
with $D_0=\bar r=1$.

For the memory case shown in Figure~5, we used
$\mu=1/2$, $D_0=\bar r=\tau=1$, and 400 logarithmically spaced points on
$10^{-3}\leq t\leq40$. For this value of $\mu$, the Prabhakar expression in
the MSD can be evaluated through an equivalent representation in terms of the
regularized lower incomplete gamma function. This representation was used
numerically because it is stable over the full time interval and avoids
cancellation in a direct summation of the three-parameter Mittag--Leffler
series.

The Fox $H$-function expressions appearing in the analytical development are
not evaluated numerically in any of the figures reported in the main text.
They are used as exact analytical representations only.

\subsection{Threshold-time calculation}

For the Markovian threshold problem, the cumulative probability inside a disk
of radius $R$ is

\begin{equation}
\Pi_{\bar r}(R,t)
=
e^{-\bar r t}\Pi_0(R,t)
+
\bar r\int_0^t
e^{-\bar r t'}\Pi_0(R,t')\,dt'.
\label{eq:threshold_num}
\end{equation}
The integral in Eq.~\eqref{eq:threshold_num} was evaluated by adaptive
quadrature. The threshold time was then obtained as the positive root of

\begin{equation}
\Pi_{\bar r}(R,t_{1/2})-\frac12=0
\end{equation}
using a bracketed Brent solver. Starting from an upper time bound equal to
unity, the bracket was enlarged by successive factors of two until a sign
change was found. The root search was restricted to $\bar r<\bar r^*$, where
$\bar r^*$ is determined from

\begin{equation}
\Pi_{\bar r}(R,\infty)-\frac12=0.
\end{equation}
For $D_0=R=1$, the resulting critical resetting rates are

\begin{equation}
\bar r^*_{\alpha=0}=6.32172,\qquad
\bar r^*_{\alpha=1/2}=4.01025,\qquad
\bar r^*_{\alpha=1}=1.92181.
\end{equation}
The curves in Figure~4 were sampled more densely close to the corresponding
critical rates in order to resolve the divergence of $t_{1/2}$.

\subsection{MFPT curves and optimal resetting rates}

The Markovian and non-Markovian MFPT curves were generated from the analytical
hitting transform derived in Appendix~A. Modified Bessel functions of the
second kind were evaluated numerically with standard double-precision special
function routines. For $D_0=R=\tau=1$ and $a=0.2$, the resetting-rate grids
contained 800 logarithmically spaced points. The intervals were
$10^{-3}\leq\bar r\leq10^2$ for the Markovian case,
$10^{-3}\leq\bar r\leq10^3$ for $\mu=3/4$, and
$10^{-3}\leq\bar r\leq10^5$ for $\mu=1/2$.

The optimal resetting rate was obtained by minimizing
$\log\langle T_{\mu,\bar r}\rangle$ as a function of $\log\bar r$ on the same
intervals, using a bounded one-dimensional minimization with tolerance
$10^{-10}$ in the logarithmic variable. The resulting values are summarized
in Table~\ref{tab:mfpt_optima}.

\begin{table}[ht]
\centering
\caption{Numerically determined optimal resetting rates and minimum MFPTs for
$D_0=R=\tau=1$ and $a=0.2$. The Markovian case corresponds to $\mu=1$.}
\label{tab:mfpt_optima}
\begin{tabular}{cccc}
\hline
$\mu$ & $\alpha$ & $\bar r_{\rm opt}$ & $\langle T\rangle_{\min}$ \\
\hline
$1$   & $0$   & $12.9891$   & $1.37788$ \\
$1$   & $1/2$ & $11.8048$   & $0.644636$ \\
$1$   & $1$   & $11.0227$   & $0.355770$ \\
$3/4$ & $0$   & $76.8127$   & $0.518644$ \\
$3/4$ & $1/2$ & $74.3507$   & $0.246988$ \\
$3/4$ & $1$   & $75.7688$   & $0.137090$ \\
$1/2$ & $0$   & $4074.39$   & $0.0427970$ \\
$1/2$ & $1/2$ & $4134.92$   & $0.0204262$ \\
$1/2$ & $1$   & $4451.28$   & $0.0111058$ \\
\hline
\end{tabular}
\end{table}

\subsection{Independent finite-difference check of the first-passage transform}

As an independent validation of the modified-Bessel representation, the
backward boundary-value problem was also solved numerically without using any
special-function solution. For a fixed Laplace variable $s>0$, we considered

\begin{equation}
\frac{D_0}{\rho^2}u''(\rho)
+
\frac{D_0(1-2\alpha)}{\rho^3}u'(\rho)
-
s u(\rho)
=
0,
\qquad
u(a)=1,
\qquad
u(L)=0.
\label{eq:fd_backward}
\end{equation}
Centred second-order finite differences reduce
Eq.~\eqref{eq:fd_backward} to a tri-diagonal linear system. We used
$a=0.2$, $R=D_0=1$, $L=6$, and grids with 1200 and 2400 intervals.
The numerical value of $u(R)$ was compared with the analytical Bessel
expression for

\begin{equation}
\alpha=0,\frac12,1,\qquad
\mu=1,\frac34,\frac12,\qquad
\bar r=1,10,100.
\end{equation}
For the memory cases the effective Laplace variable was

\begin{equation}
s_{\rm eff}
=
\frac{(\bar r\tau)^\mu}{\tau}.
\end{equation}
Across these 27 parameter combinations, the largest relative discrepancy
between the 2400-interval finite-difference result and the analytical hitting
transform was $4.46\times10^{-5}$. This provides an independent numerical
check of the backward first-passage calculation and of the subordination
replacement used in the memory case.

\subsection{Software and reproducibility}

The calculations were implemented in Python 3 using NumPy, SciPy, Matplotlib,
and mpmath. In particular, the numerical implementation uses routines for
modified Bessel functions, incomplete gamma functions, adaptive quadrature,
Brent root finding, and bounded scalar minimization. The data used in the
figures are exported as two-column text files containing the horizontal
variable and the corresponding observable.


\end{document}